\documentclass[]{aa} \usepackage{psfig} \usepackage{ifthen}
\begin{document}
\title{Gravitational  microlensing of  stars with  transiting planets}

\author{Geraint F.  Lewis}

\offprints{G.  F.  Lewis}

\institute{ Anglo-Australian  Observatory, P.O.  Box  296, Epping, NSW
1710, Australia: Email \tt{gfl@aaoepp.aao.gov.au}}

\date{Received January 2001; accepted October 2001}

\abstract{If planetary systems are ubiquitous then a fraction of stars
should possess a transiting planet when being microlensed.  This paper
presents  a study  of the  influence of  such planets  on microlensing
light  curves.   For  the   giant  planets  recently  identified,  the
deviations  in  the  light  curve  can be  substantial,  although  the
specifics of  the perturbations are  dependent upon the radius  of the
planet relative to  that of the star, the location  of the planet over
the  stellar surface  and  the orientation  of  the sweeping  caustic.
Given  that the  instantaneous probability  of  transiting hot-Jupiter
like planets is  small, less than a percent, and  only a proportion of
microlensing  events  {  exhibit   caustic  crossing  events},  {  the
probability  of detecting  a transiting  planet during  a microlensing
event  is   small,  $\sim10^{-6}$.   However,  a   number  of  factors
influencing this  probability, such as  the number of solar  type star
that  possess planets, are  uncertain, and  the prospect  of detecting
transiting planets  in future  large catalogues of  microlensing light
curves may be viable.}  The results of this study also have bearing on
the   gravitational    microlensing   of   spots    on   the   stellar
surface. }

\maketitle
%

\section{Introduction}\label{introduction}
More than  fifty planets  have now been  identified outside  the Solar
system  (see {\tt http://exoplanets.org}  for a  summary of  the known
planetary systems).  While  the majority of these have  been found via
the  measurement of  a  stellar  radial velocity  changes  due to  the
presence of an  orbiting planet, more recently the  dimming of stellar
light  due to  the transit  of  a giant  planet has  been observed,  {
specifically in the case of  HD209458} (Charbonneau et al. 2000; Henry
et al.  2000;  Queloz et al.  2000; Castellano et  al.  2000; Doyle et
al.  2000;  Gaudi 2000).  Typically, light  curves are seen  to dip by
$1-2\%$ as  the planet transits  the star, corresponding  to planetary
radii of $0.10-0.14\ R_s$.   Given current observational limits, these
techniques can only  be used to identify planets within  a few tens of
parsecs.

Gravitational microlensing has the  potential to discover planets over
a much  greater distance.  Perturbing  the magnification distribution,
and hence  the form  of the microlensing  light curve, planets  can be
identified  orbiting compact objects  in the  Galactic Halo  (Mao {\&}
Paczy{\'n}ski  1991; Gould {\&}  Loeb 1992;  Bolatto {\&}  Falco 1994;
Bennett  {\&} Rhie  1996; Wambsganss  1997). More  recently,  Graff \&
Gaudi (2000) and Lewis \& Ibata (2000) turned their attention to using
gravitational microlensing  to identify planets orbiting  stars in the
Galactic   Bulge,    that   is   detecting    planets   orbiting   the
\underline{source star} in a  microlensing event, rather than orbiting
the lens.   Assuming that the  star plus planet  system is swept  by a
fold caustic  formed by a  binary lens, it  was shown that  the feeble
light   that   is   reflected   from   the  planet   ${\rm   (up~   to
\sim10^{-4}L_*)}$ can be magnified  to observable levels, resulting in
a $\delta  M\sim0.03$mag fluctuation.  The  scattering responsible for
reflecting the stellar light from  the planet also acts to polarize it
and  Lewis \&  Ibata (2000)  demonstrated that  this signature  too is
boosted to  observable levels, probing the physical  conditions in the
planetary  atmosphere.   Following these  initial  studies, Ashton  \&
Lewis (2001)  considered the influence  of the planetary phase  on the
form of  the microlensing  light curve, finding  that when  the planet
appears crescent-like, the  magnification can be substantially greater
than the simple, circularly symmetric model that was employed in these
earlier studies. {  While the resulting flux is  still lower than that
of a  microlensed planet at opposition,  this additional magnification
does aid detection of crescent-like planets.}

This  paper also  focuses on  the identification  of  planets orbiting
stars   that  are  the   sources  during   gravitational  microlensing
events. Instead of looking at the reflected light, however, this paper
examines { what effect a planet  transiting a stellar surface has on a
microlensing  light curve.  Section~\ref{method}  presents simulations
of such microlensing situations, considering first a simple model with
a    star    and     static    planet    (Sect.~\ref{ideal}),    while
Sect.~\ref{realsims} this  study is expanded to include  the effects {
of  planetary  motion.}    }   Section~\ref{discussion}  compares  the
results of this study with previous investigations of the influence of
spots on the  stellar surface, as well as  investigating the frequency
of     planetary      transits     during     microlensing     events.
Section~\ref{conclusions} presents the conclusions of this study.

\begin{figure}
{
\centerline{ \psfig{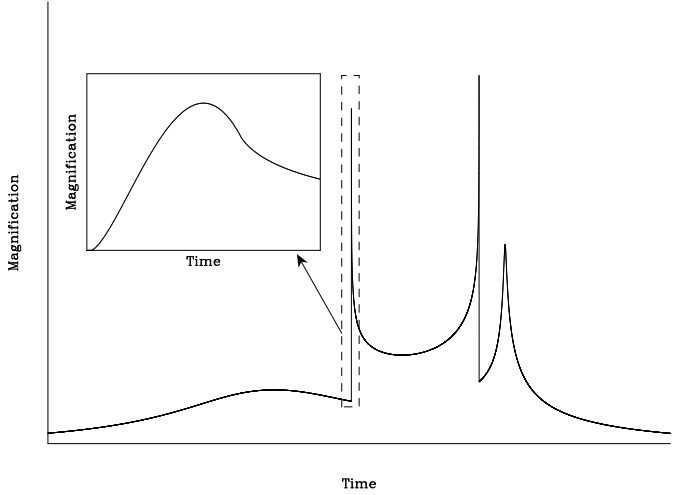} }}{}
\caption{{  A  { sample  binary  caustic-crossing} microlensing  light
curve,  showing the very  sharp features  characteristic of  a caustic
crossing.  This paper concerns itself only with the high magnification
regions of the  light curve, where the caustic  actually sweeps across
the source.  This  is highlighted in the inset box.  The time scale of
the microlensing event is discussed in Sect.~\ref{realsims}.}}
\label{fig1}
\end{figure}

\section{Simulations}\label{method}
\subsection{Microlensing}\label{microlensing}
A  wealth of information  has resulted  from the  rapid growth  in the
field of Galactic and  Local Group gravitational microlensing over the
last decade.   The reader is  directed to the review  of Paczy{\'n}ski
(1996) which covers this background material in detail.

\begin{figure*}
{
\centerline{ \psfig{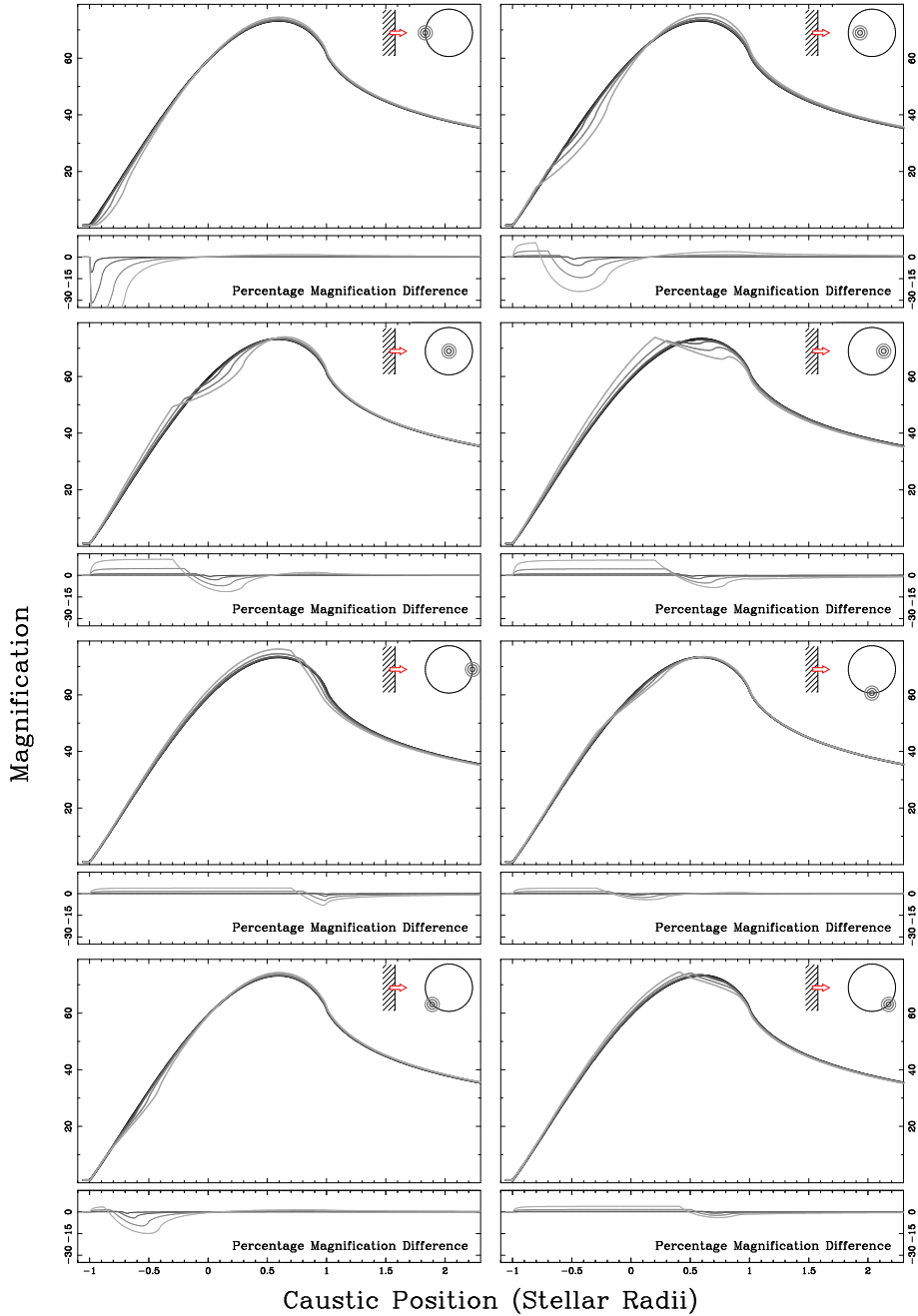} }}{}
\caption{The   microlensing    light   curves   for    the   planetary
configurations  described in the  text.  In  each panel,  the relative
position of the star, caustic  (hashed region being the region of high
magnification, with  the arrow denoting  its direction of  travel) and
planets are  presented in the top  right hand corner.   Each series of
light curves in a particular panel  are for a range of planetary radii
{ (0.035, 0.1, 0.2 \& 0.3  $R_s$)}, the greyscale for each light curve
corresponding to the  greyscale of the planet in  the upper portion of
each figure.   These light curves  should be compared to  the exampled
light curve given in Fig.~\ref{fig1}.  The lower section of each panel
presents the percentage difference in the magnification between a star
with a  transiting planets  and the same  star without such  a planet.
The  units on  the  abscissa denote  the  position of  the caustic  in
stellar radii, relative to the centre  of the star at the origin.  The
total observed duration of such caustic crossings is $\sim$10 hours.}
\label{fig2}
\end{figure*}

Most  studies   of  Galactic  halo  microlensing   have  focused  upon
microlensing  by  isolated  compact  objects,  which  display  simple,
bell-shaped  light curves.   A  fraction of  events, however,  display
several  rapid,  asymmetric   features  which  are  characteristic  of
microlensing by  a binary  system (Alcock et  al.  2000),  { presented
schematically in  Fig.~\ref{fig1}.}  These features are  the result of
the  presence of extended  caustics which  are associated  with binary
lenses.  The  large magnifications which result when  a caustic sweeps
across a  source has proved  to be a  powerful diagnostic of  not only
stellar  systems   (Agol  1996;  Han,  Park,  Kim   {\&}  Chang  2000;
Heyrovsk\'{y},  Sasselov {\&} Loeb  2000), but  also, at  high optical
depths, structure at the heart of quasars (Wambsganss \& Paczy\'{n}ski
1991;  Lewis \&  Belle 1998;  Agol {\&}  Krolik 1999;  Belle  \& Lewis
2000).   Caustics   in  such  networks  are  comprised   of  `{  fold}
catastrophes' (Schneider, Ehlers \&  Falco 1992), combining in regions
to  form higher  order  catastrophes.  As  they  dominate the  caustic
structure  formed by a  binary lens  in the  following analysis  it is
assumed that the  source star and planet are swept  by a fold caustic.
It is  assumed that the caustic is  straight in the vicinity  of the {
source star  and} planet. {  It should be  noted, however, that  for a
small number of microlensing  light curves the observed variations are
are  consistent  with  the  source  star being  swept  with  cusp-like
caustics  (see  Alcock  et   al.   2000).   Due  to  their  curvature,
microlensing events due to cusp caustics can result in different light
curves (Fluke \& Webster 1999).}

As a point  source is swept by a fold caustic,  the magnification at a
location $x$ is given by
\begin{equation}
\mu(x) = \sqrt{\frac{g}{x-x_c}}H(x-x_c) + \mu_o\ ,
\label{caustic}
\end{equation}
where  $\sqrt{g}$ is  the ``strength''  of the  caustic, $x_c$  is the
location of the caustic and  $H(x)$ is the { Heaviside} step function.
The background magnification, $\mu_o$,  is due to other lensed images,
not  associated  with  the  caustic  crossing, formed  by  the  binary
lens. In the following study it is assumed that $\mu_o=1$.

An examination  of Equation~\ref{caustic} reveals that  as the caustic
crosses the point source, where $x-x_c=0$, the resulting magnification
is  infinite.   With  any  finite  source,  however,  integrating  the
magnification  distribution  over  the  source  results  in  a  finite
magnification,  smoothing  out the  microlensing  light  curve in  the
vicinity of the  caustic (see inset box in  Fig.~\ref{fig1}). The peak
magnification in the light curve for  a source swept by a fold caustic
is given by
\begin{equation}
\mu_{peak} = f \sqrt{\frac{g}{R_s}}
\label{peakmag}  
\end{equation}
where $R_s$ is the radius of the source, and $f$ is a form factor that
accounts for the  specific geometry of the source  (Chang 1984). For a
uniformly {  bright} circular source,  $f=1.39$.  With regards  to the
work presented in Sect.~\ref{ideal},  these parameters are chosen such
that $\mu_{peak}\sim70$,  a value typical  for microlensed stars  of a
Solar radius in the Galactic Bulge (Lewis \& Ibata 2000).  The results
in this paper  are simply scalable to other  stellar radii and caustic
strengths using Equation~\ref{peakmag}.

\subsection{Idealized Simulations}\label{ideal}
{  The magnification  of a  star during  a  gravitational microlensing
event is  a powerful probe of  the physical conditions  of the stellar
surface  (Minniti et  al. 1998;  Gaudi \&  Gould  1999; Heyrovsk\'{y},
Sasselov {\&} Loeb 2000). One particular observational success in this
regard  is  the  study  of  Albrow  et al  (1999b)  who  measured  the
limb-darkening of K  giant star, the source of  the microlensing event
97-BLG-28.  Such a measure is only possible because the details of the
surface brightness of the source star influences the form of the light
curve during a gravitational microlensing event.  As all stars will be
limb-darkened to  some degree,  resulting in slightly  different light
curves.  Hence,  limb-darkening is introduced by  modeling the surface
brightness of the star to be (Albrow et al. 1999a; Afonso et al. 2000)
\begin{equation}
S(R) = 1 - \Gamma\left[ 1 - 
\frac{3}{2}
\sqrt{\left( 1 - \frac{R^2}{R_s^2}\right)}\right]
\label{limbd}
\end{equation}
where  $\Gamma$ is  a limb-darkening  parameter, such  that $\Gamma=0$
corresponds  to  a  star  with  a uniform  surface  brightness,  while
$\Gamma=1$  the star  is centrally  bright, fading  to zero  at $R_s$.
Outside   the   range   $0\leq\Gamma\leq1$  the   surface   brightness
distribution possesses negative regions and is, therefore, unphysical.
In the following, the surface  brightness of the star is parameterized
with a limb-darkening of $\Gamma=0.3$, which is approximately the mean
value found for  the microlensing event 98-SMC-1 (Afonso  et al 2000),
although it  is found that, while  the form of  the microlensing light
curve is strongly dependent upon  the value of $\Gamma$, the magnitude
of the influence of a transiting planet is not.}

The  planet is  also represented  as a  circular disk  that completely
obscures  a  fraction  of  light  from the  star.   {  Four}  fiducial
planetary radii  were investigated, with  $R_p =$ { 0.035},  0.1, 0.2\
\&\  0.3$R_s$.  With  regards  to  our own  Solar  system, Jupiter  is
$\sim0.1R_s$, while Uranus and Neptune are $\sim0.036R_s$; these would
induce  a   transit  dimming  of   the  Sun  of  $1\%$   and  $0.13\%$
respectively. {For these initial  simulations,} it is assumed that the
time  scale for  the caustic  crossing  the stellar  surface is  small
compared to the  orbital motion of the planet can  be neglected. { The
orbital motion of the planet is considered in Sect.~\ref{realsims}.}

For a source with an  arbitrary brightness profile, the calculation of
its light curve as it is swept by a caustic is typically obtained with
a brute-force integration (see Ashton \& Lewis 2001).  { Therefore, to
determine the magnification, such  a brute-force approach was adopted,
numerically  integrating   Equation~1,  weighted  by   the  brightness
distribution over  the disk of the  star, and dividing by  the flux of
the star  in the absence  of the lens  [see Lewis \& Belle  (1998) and
Ashton  \& Lewis  (2001)  for  a description  of  the approach].   The
accuracy of  the numerical integration was ensured  with comparison to
analytic results.}

Figure~\ref{fig2}   present  the   magnifications   and  light   curve
deviations for the models discussed  in the previous section.  In each
panel,  the schematic in  the upper  right-hand corner  represents the
relative  orientation of  the  star, planet  and  caustic. The  hashed
region denotes  the region of  high magnification associated  with the
caustic, and the arrow its  direction of motion.  The upper section of
each panel presents the microlensing  light curve in the region of the
peak  magnification.  {  The  flux  received by  an  observer is  this
magnification multiplied by  the unlensed flux of the  source, a point
returned to in Sect.~\ref{realsims}.}  The lower regions of each panel
presents the  percentage difference  in the observed  { magnification}
between  the light  curves of  a star  with and  without  a transiting
planet;
\begin{equation}
\delta \mu_\% = 100.\left( \frac{{ \mu}_T}{{ \mu}} - 1
\right)
\end{equation}
The scale on the abscissa on all panels is the location of the caustic
with respect to  the centre of the star, which  resides at the origin.
{  The  time  scale  of  these microlensing  events  is  discussed  in
Sect.~\ref{realsims}.}

In  the various  panels of  Fig.~\ref{fig2} the  planet is  located at
several differing  positions over the stellar surface.   Some of these
lie completely over the star, while with others only a fraction of the
planet obscures the star.  It is immediately apparent that the form of
the microlensing  light curve is  very dependent upon the  position of
the planet over the stellar surface relative to the orientation of the
sweeping  caustic.   If  the  planet  is  located  where  the  caustic
initially  crosses onto  the star,  substantial  fractional deviations
from the {  case without a planet can be}  seen, the largest deviation
occurring for  the largest planets (a  feature that is seen  in all of
the light curves presented in this paper).  For planets placed further
across the  stellar surface,  deviations can be  seen to occur  in the
light  curve, although the  location of  these deviations  is strongly
dependent  upon the planetary  location.  When  the planet  is located
nearer to the far side of the stellar surface, the deviations from the
uniformly  { bright}  star become  quite dramatic  in vicinity  of the
light curve peak, displaying  quite complex structure, with deviations
of  up to  $\sim10-20\%$ for  the largest  planets.  For  the smallest
planets considered, however, the fractional deviation is comparatively
small, typically peaking at a couple of percent.

{
\subsection{Realistic Simulations}\label{realsims}
{  In the previous  section, it  was assumed  that the  planet remains
fixed  over the  surface  of  the source  star  as it  is  swept my  a
microlensing  caustic.   Here, this  assumption  is  examined in  more
detail.}

\begin{figure*}
{                               \centerline{
\psfig{figure=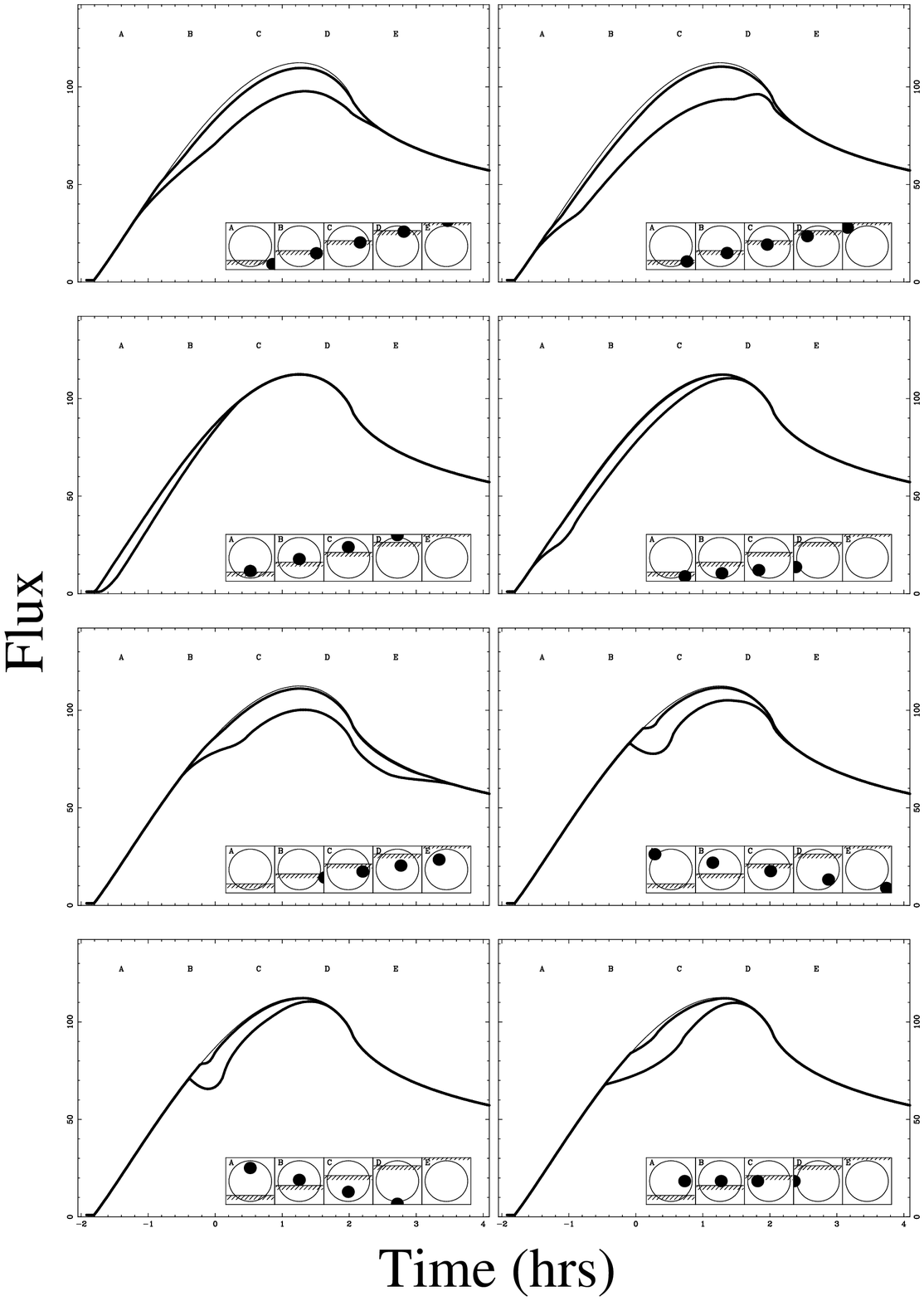,height=19.cm,angle=0.0} }}{}
\caption{Several  examples   of  realistic  microlensing  simulations,
presenting  flux verses time  for several  planetary paths.   The thin
lines correspond to the microlensing light curves without a transiting
planet, while those  possessing a small deviation from  this curve are
for a  planet of radius 0.1$R_s$.  { The large  deviation light curves
are for a  planet with a radius of 0.3$R_s$.}   Each panel possesses a
series of  boxes which illustrate the  position of the  planet and the
caustic over the stellar surface.   The high magnification side of the
caustic is  denoted by  a hashed region.   Each box is  labeled A...E,
these correspond to  the points on the light curves  below each of the
letters in the main panel.}
\label{fig3}
\end{figure*}

An  examination of  a catalogue  of binary  microlensing  light curves
(Alcock et al.  2000) reveals  that the total microlensing event (such
as that  portrayed in Fig.~1.)   can span $\sim$20 to  several hundred
days.  The time scale for the caustic sweeping across the source star,
as denoted in the inset box of Fig.~1., is much smaller. Examining two
cases in the literature, the light curves presented in Fig.~\ref{fig2}
represent $\sim$17 hours in the case of 98-SMC-1 (Afonso et al.  2000)
and $\sim$7.5  hours for  96-BLG-3 (Alcock et  al.  2000), {  with the
time for the caustic to  cross the stellar surface being $\sim9$ hours
and  $\sim4$ hours  respectively}.  It  is the  value for  this latter
event  that will  be adopted  for  the simulations  presented in  this
section.

How  does  this  time scale  compare  to  the  transiting time  of  an
extrasolar planet? Examining the  light curve of HD209458 (Charbonneau
et al. 2000)  reveals that the planet { to  cover the apparent stellar
diameter is $\sim$2.3 hours (this is slightly shorter than the transit
time {  if the  HD209458's planet is  orbiting with an  inclination of
90$^o$})},  comparable to  the caustic  sweeping time.   Therefore, in
fully  simulating  the  influence  of  a transiting  planet  during  a
microlensing event  requires changing the position of  the planet over
the  stellar surface during  the caustic  crossing.  In  the following
study, the transiting time scale for HD209458 will be employed.

Figure~\ref{fig3}  presents  eight  examples  of the  microlensing  of
planetary transits. Each panel possesses a number of curves.  The thin
curves present the  light curves in the case  of no transiting planet.
Two planetary radii are considered  in this plot, represented as thick
solid lines,  one of 0.1$R_s$,  the other of  0.3$R_s$; distinguishing
these  two is  simple as  the smaller  planet always  produces smaller
deviations  from the  case  with  no transiting  planet.   { Again,  a
limb-darkening parameter of $\Gamma=0.3$ is employed.  Each panel also
presents a  series of boxes,  displaying the relative position  of the
planet  (dark circle)  over the  stellar surface  (open  circle).  The
caustic, denoted  by a line with  a hashed region  indicating the high
magnification side,  is also  presented.  Each box  is labeled  with a
letter, A..E; these  correspond to the times along  the light curve as
indicated by  the lettering  at the top  of each panel.   An important
point to note is that each light curve present flux verses time and so
takes into account the slight dimming  that occurs when a planet is in
front of the  stellar surface, which, for a  planet of radius 0.3$R_s$
corresponds to a dip of almost $\sim10\%$.}

In examining the light curves presented here, there are several points
to note.   Firstly, as seem  in the previous sections,  larger planets
have a larger impact on  the microlensing light curves. Secondly, if a
planet is located  over the stellar surface but in a  region yet to be
swept by the caustic it produces no effect other than a slight dimming
of several percent of the {\it unlensed} source flux; { this change is
negligible  compared to  the flux  form the  magnified portion  of the
star.  Of course, when the planet has moved completely off the stellar
surface it has no influence on  the light curve which becomes the same
as case without a transiting planet.

It  is also  apparent that  the path  of the  planet over  the stellar
surface, relative to the sweeping caustic, greatly influences the form
of the resultant  light curve, with some light  curves for a planetary
radius of 0.3$R_s$, displaying variations of up to $\sim$20\% from the
situation with  no transiting  planet.  For both  Figs.~\ref{fig2} and
\ref{fig3}  it is  important to  emphasis that  the deviations  in the
light  curves possess  quite  characteristic forms  and  would not  be
confused with an incorrectly assumed limb-darkening parameter etc. For
large  planets,  such  deviations   would  be  apparent  in  the  high
signal-to-noise,  fine  temporal   sampling  light  curve  of  caustic
crossings already obtained (Afonso et al. 2000).}

\section{Discussion}\label{discussion}
\subsection{Transit Frequency}\label{frequency}
Assuming randomly oriented {  circular} orbits, the probability that a
planet  of  radius  ${R_p}$  will  be  found {  at  a  given  instant}
transiting the surface of a star of radius ${R_s}$ is
\begin{equation}
P_T = \frac{1}{2} \left(1-\sqrt{1-\left(\frac{
R_s + R_p}{a}\right)^2}\right) 
\label{transitprob}
\end{equation}
where ${a}$ is the radius of the orbit. As an example, we consider the
parameters of the recently  identified transiting planet of HD~209458,
a G0 star with a radius  of 1.1$R_\odot$ (Henry, Marcy, Butler \& Vogt
2000;  Charbonneau,  Brown, Latham  \&  Mayor  2000).   The planet  is
orbiting at 0.046AU and possesses a radius of $\sim1.3R_J$, suggesting
a transit probability of $P_T\sim0.4\%$.  As $R_p/R_s\sim 0.12$, the {
planet would cause deviations of several percent in a caustic crossing
microlensing event} (see Fig.~\ref{fig3}).

{ In calculating the expected rate of planetary transit events seen in
microlensing light curves,  a number of points need  to be considered.
Firstly,  what  proportion  of  microlensing  events  possess  caustic
crossing   events;  Di~Stefano   (2000)   considered  this   question,
concluding  that  if  the  entire  halo  population  was  composed  of
binaries,  then $\sim6\%$ should  present caustic  crossings, although
this may  slightly underestimate, due to observational  bias, the true
rate (Alcock  et al. 2000). This  is, of course, modified  by the halo
fraction of binary MACHOs, $f_{bin}$,  which has yet to be determined.
Also  important  is  the   fraction  of  microlensed  stars  that  are
$\sim1R_\odot$ in  size; as it  is the relative  size of the  star and
planet that is important, the transit of even a hot (close-in) Jupiter
around  a giant  star will  not result  in a  significant  light curve
deviation. { Alcock et al. (2000) found that $\sim66\%$ of stars whose
size  could  be determined  in  caustic  crossing microlensing  events
possessed    radii    $\leq    1.5R_\odot$,   the    smallest    being
$\sim0.9R_\odot$; this value is  adopted here.}  Finally, there is the
proportion of solar type stars that possess a hot Jupiter-like planet;
again,  given  our  current  observations,  this number  if  not  well
determined, although  recent studies suggest that this  is $\sim3\%$ {
for planets orbiting at less than $\sim0.1$AU} (Cumming et al.  1999).
Combining these, the fraction of microlensing events that are expected
to show evidence of transiting planets is $\sim5\times10^{-6}f_{bin}$.
While  this indicates  that the  identification of  transiting planets
during microlensing  events is a very rare  occurrence, several factor
are uncertainty  and future revisions  may make the  identification of
transiting   planets  in  microlensing   surveys  a   more  attractive
prospect.}

\subsection{Stellar Spots}\label{stellarspots}
The dimming produced  by a transiting planet mimics  a completely dark
spot  on  the  stellar  surface.   Heyrovsk{\'y}  \&  Sasselov  (2000)
considered  the microlensing  of a  spotty star  by a  single isolated
mass.  In  general, this resulted  in only small perturbations  to the
microlensing  light curve,  although larger  deviations  were possible
with  the alignment of  the spot  with the  point-like caustic  of the
isolated lens.  Han,  Kim, Chang \& Park (2000)  extended this earlier
work and considered a spotty  star microlensed by the extended caustic
structure of a binary lens, akin to the model presented in this paper.
{ Han, Kim, Chang  \& Park (2000) placed a spot with  a contrast of 10
projected solely at the centre  of the star and assuming that changing
the location of  the spot does not greatly  affect the resultant light
curve.  Considering  the work presented  in this paper,  however, this
latter assumption turns  out not to be true,  as changing the location
of  the spot,  like changing  the location  of the  planet, introduces
quite different  deviations into the  light curve. Hence, the  work of
Han, Kim, Chang  \& Park (2000) needs to be  extended to consider this
effect {  (their study also  requires the addition of  a limb-darkened
stellar surface brightness profile to obtain `realistic' results.)} }

While it appears that {  the microlensing signature of a} stellar spot
can imitate  the microlensing signature of a  transiting planet, there
are several features of each  that can ease the differentiation of the
two phenomena. Firstly,  while a star with a single  large spot can be
envisaged,  { many  are likely  to present  a number  of spots  on the
stellar  surface.}  This would  result in  multiple variations  in the
microlensing  light  curve,  as  opposed  to the  variations  seen  in
Fig.~\ref{fig2}.  Multiple planets  transiting the stellar surface may
mimic such a spotty  star, but considering the probabilities presented
in Sect.~\ref{frequency}, such occurrences  will be extremely rare.  {
Unlike   planets,  however,  spots   are  not   necessarily  circular,
presenting possibly peculiar forms  to the sweeping caustic.  Circular
spots also  appear elliptical when at  the limb of a  star, leading to
deviations from the expected  planetary light curves presented in this
paper.   As noted  previously, stellar  spots need  not  be completely
black  against  the  stellar   surface,  possessing  both  colour  and
intensity  structure.  These would  introduce {  additional} chromatic
and   spectroscopic   variations   to   the  microlensing   {   event}
(Heyrovsk\'{y}   and   Sasselov   2000;   Bryce  and   Hendry   2000).
Unfortunately,  the identification  of such  spectroscopic variability
does not  cleanly differentiate between the two  as transiting planets
in themselves can introduce such features (Queloz et al.  2000; Jha et
al.  2000), as  can microlensing of a rotating  stellar surface (Gould
1997).   Hence, a  more  careful spectroscopic  analysis, taking  into
account the various sources  of spectroscopic variability, is required
before  a conclusive  identification  of a  transiting  planet can  be
made.}  { To provide the framework  for such an analysis, and to fully
determine  the  photometric   and  spectroscopic  differences  of  the
microlensing signature  of a  spotty star and  a transiting  planet, a
more extensive numerical study of each case is required.}

\section{Conclusions}\label{conclusions}
Several  studies  have  demonstrated  that  planets  orbiting  compact
objects  in the Galactic  disk/halo can  perturb the  form of  a light
curve during  a microlensing  event.  More recently,  researchers have
turned their attention to using microlensing to identify planets orbit
the source  stars in  the Galactic bulge  and Magellanic  Clouds.  The
resulting deviations in  the form of the microlensing  light curve are
small  and   are  observationally  tricky   with  today's  technology.
However, a  number of specialized monitoring teams  (MOA, PLANET, GMAN
etc) have  succeeded in  collecting many hundreds  of data  points for
several  microlensing  events,  with  high photometric  accuracy,  and
demonstrate  that  it  is  possible  to detect  such  small  amplitude
deviations.

This  paper has investigated  this scenario  further by  examining the
microlensing  of   a  star   which  possesses  a   transiting  planet.
Jupiter-like  planets transiting  Sun-like stars  produce  a $\sim1\%$
dimming of the star light.   If such a system is microlensed, however,
the  presence  of  a planet  transiting  the  star  can lead  to  very
characteristic deviations to the form of the microlensing light curve.
The degree of  the deviation is very dependent upon  the radius of the
planet and  the relative position of  the planet in front  of the star
and the orientation of the caustic crossing.

In flux, the light curve  profile can deviate strongly ($\sim20\%$ for
the  largest planets  considered), with  quite  characteristic shapes,
from  that  expected  from   the  microlensing  of  star  without  the
transiting   planet,    making   identification   possible.    Smaller
deviations,  corresponding to  relatively smaller  transiting planets,
can be  found by comparing  separate high magnification  events during
binary  microlensing  events   (see  Fig.~\ref{fig1}),  as  these  are
separated by  several days and the  planet will have moved  due to its
orbital motion to another location  over the stars surface, or, { more
probably, completely} off the stellar disk.  As the characteristics of
the  caustic crossing  will also  be different,  in both  strength and
sweeping  direction, the  resulting light  curve of  this  second high
magnification event  will be  different, even if  the location  of the
planet over the  stellar surface does not change.   The details of the
light  curve  reveal  the   underlying  caustic  structure  of  binary
microlens (Alcock  et al 2000) and,  for events that do  not possess a
transiting  planet,  the  surface  brightness  profile  of  the  star,
including the  effects of limb-darkening, can be  determined in detail
(Albrow et  al 1999b).  This information  can be used  to untangle the
influence of  the stellar profile  during the microlensing  event with
the transiting planet.

{ The  deviations formed  by the presence  of a transiting  planet can
mimic the effects  of a spot on the stellar  surface. For many systems
differentiating between planets and spots will be straight forward, as
a  star can  possess multiple  spots, where  only a  single transiting
planet is expected.  Other clues  come from the non-circular nature of
spots and the fact that  spots can possess strong colour and intensity
variations   across  them,  resulting   {  in   additional}  chromatic
deviations over the light curve, { although a number of other spectral
signatures as a  result of microlensing of the  stellar surface or the
presence  of the transit  planet will  make this  differentiation more
difficult.  Finally, given  the  rarity of  hot  Jupiter planets,  the
possibility of transits  and the low binary microlensing  rates in the
Galaxy,  the   prospects  for  detecting   planetary  transits  during
microlensing  are  poor.   Some  of  the parameters  {  necessary  for
determining}  the expected  rate  are  poorly known  and  so a  future
revised estimate may make the proposition more favourable. The results
in  this study,  however, are  relevant  to the  study of  the more  {
frequently expected detection of} stellar spots during microlensing.}
 
\section{Acknowledgments}
Joachim   Wambsganss  is   thanked   for  enlightening   conversations
gravitational  microlensing  and   extrasolar  planets,  and  for  his
comments on  a previous  draft of  this paper, {  and Chris  Tinney is
thanked for  discussions on planetary systems.  The  referee, { David}
Heyrovsk{\'y}, is thanked for constructive comments.}

\end{document}